\documentclass[twoside]{dis09}
\usepackage[latin1]{inputenc}
\usepackage[dvips]{graphicx,epsfig,color}
\usepackage{wrapfig,rotating}
\usepackage{amssymb,amsmath,array}

\begin{document}
\title{The Color-Dipole Picture and $F_L$}

\author{Dieter Schildknecht 
%
\thanks{Supported by Deutsche Forschungsgemeinschaft, contract 
number Schi189/6-2.}
\thanks{Presented at DIS 2009, 
Madrid, April 25 to 30, 2009.}
%
\vspace{.3cm}\\
%
Fakult\"at f\"ur Physik, Universit\"at Bielefeld, D-33615 Bielefeld, Germany  \\
and \\
Max-Planck Institut f\"ur Physik (Werner-Heisenberg-Institut), D-80805
M\"unchen, Germany
%
}

\maketitle

\begin{abstract}
The prediction of $F_L (x , Q^2) = 0.27 F_2 (x, Q^2)$ in the color-dipole
picture, based on color-transparency and transverse-size reduction, is 
consistent with the experimental results from HERA.
\end{abstract}


We consider the photon-nucleon interaction at low $x_{bj} \cong 
Q^2/W^2 \ll 0.1$, such that 
\begin{equation}
\frac{1}{\Delta E} = \frac{1}{x_{bj} + \frac{M^2_{q \bar q}}{W^2}}
\frac{1}{M_p} \gg \frac{1}{M_p} . 
\label{1}
\end{equation}
The covariant quantity in (\ref{1}) is identical to the life-time of a 
hadronic $q \bar q$ fluctuation of mass $M_{q \bar q}$ of the photon 
in the rest frame of the nucleon. The inequality (\ref{1}) is the 
space-time condition \cite{1} for the validity of generalized vector dominance
\cite{2}. 

The $\gamma^* p$ scattering process at low $x_{bj}$ proceeds via 
$q \bar q$ scattering. The $q \bar q$ state interacts via gluons coupled
to both the quark and the antiquark, i.e. it interacts as a 
color-dipole state (color-dipole picture, CDP) \cite{3}. 

\begin{wrapfigure}{l}{0.5\columnwidth}
\vspace{-12pt}
\centerline{\includegraphics[width=0.45\columnwidth]
{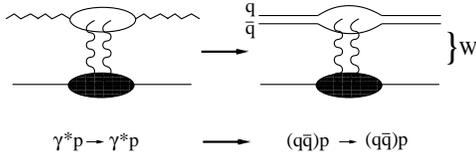}}
\vspace{-12pt}
\caption{The color-dipole interaction.}\label{Fig:schi-gluon}
\end{wrapfigure}
The mass of a $q \bar q$ fluctuation, $M_{q \bar q}$, is restricted by
\begin{equation}
m^2_{\rho^0} \le M^2_{q \bar q} \le m^2_1 (W^2) , 
\label{2}
\end{equation}
where $m^2_1 (W^2) \ll W^2$ approximately coincides with the upper 
end of the diffractive mass spectrum observed at HERA. The frequently 
adopted approximation of $m^2_1 (W^2) \rightarrow \infty$ restricts the 
kinematic domain of validity of the CDP.

Consider a timelike photon of mass $M_{q \bar q}$. The structure of its 
$\gamma^* (q \bar q)$ coupling implies an enhancement \cite{4} of the transverse
size of the $(q \bar q)^{J=1}_T$ state of mass $M_{q \bar q}$ and 
spin $J=1$ originating from a transversely polarized photon, relative to 
the transverse size of the $(q \bar q)^{J=1}_L$ state originating from a 
longitudinally polarized photon. The transverse-size enhancement implies 
an enhanced cross section, 
\begin{equation}
\sigma_{(q \bar q)^{J=1}_T p} (M^2_{q \bar q}, W^2) = \rho \sigma_
{(q \bar q)^{J=1}_L p} (M^2_{q \bar q} , W^2), 
\label{3}
\end{equation}
where \cite{4}
\begin{equation}
\rho = \frac{4}{3} .
\label{4}
\end{equation}
The factor $\rho$ is independent of the Lorentz boost from the 
$(q \bar q)$ rest frame to the energy $W$ of the $(q \bar q)p$ 
interaction; $\rho$ is independent of $W$. 

The transition from the interaction of a timelike photon, 
$\gamma^*_{L,T}$, of mass $M_{q \bar q}$ to the interaction of a 
spacelike one of four momentum squared $q^2=-Q^2 < 0$ in the 
imaginary part of the forward Compton amplitude requires integration
over all masses of the incoming and outgoing $q \bar q$ fluctuations. 
Upon introducing the transverse size, $r_\bot$, of a $q \bar q$
fluctuation and upon introducing the $(q \bar q)^{J=1}_{L,T} p$
scattering cross section for spin $J=1$ quark-antiquark dipole states, the 
photoabsorption cross section in the CDP becomes \cite{5,4}
\begin{equation}
\sigma_{\gamma^*_{L,T}p} (W^2 , Q^2) = 
\frac{2\alpha R_{e^+e^-}}{3\pi^2} Q^2 \int d^2 r^\prime_\bot K^2_{0,1}
(r^\prime_\bot Q) \sigma_{(q \bar q)^{J=1}_{L,T}p} (r^\prime_\bot , 
W^2).
\label{5}
\end{equation}  
Massless quarks are assumed in (\ref{5}). 
Quark masses can be introduced via quark-hadron duality. The 
variable $r^\prime_\bot$ is related to the transverse size of a $q \bar q$
state via $r^\prime_\bot = r_\bot \sqrt{z(1-z)}$ where 
$0 \le z \le 1$, and $R_{e^+ e^-} = 3 \sum Q^2_q$, where $Q_q$ denotes 
the quark charge, and $Q \equiv \sqrt{Q^2}$. 
The representation (\ref{5}) of the CDP factorizes the $\gamma^*_{L,T} p$ 
cross section into the $Q^2$-dependent (square of the) photon wave function,
given by the modified Bessel function $K_{0,1}(r^\prime_\bot Q)$, and the 
$W$-dependent dipole cross section $\sigma_{(q \bar q)^{J=1}_{L,T}} (r^\prime_
\bot , W^2)$. Since $\gamma^* p$ interactions at low $x_{bj}$ proceed 
via (on-shell) $q \bar q$ scattering, the frequently employed 
factorization in $(Q^2, x_{bj})$ rather than in $(Q^2, W^2)$ can at most 
be of approximate validity \cite{6}. The transverse-size enhancement (\ref{3})
enters (\ref{5}) via 
\begin{equation}
\sigma_{(q \bar q)^{J=1}_T p} (r^\prime_\bot , W^2) = \rho \sigma_
{(q \bar q)^{J=1}_L p} (r^\prime_\bot , W^2).
\label{6}
\end{equation} 

The interaction of the $q \bar q$ state as a color-dipole state 
requires a representation of the dipole cross section in (\ref{5}) 
given by \cite{3,5,4}
\begin{eqnarray}
\sigma_{(q \bar q)^{J=1}_{L,T}}  (r^\prime_\bot , W^2) && =
\int d^2 l^\prime_\bot \bar{\sigma}_{(q \bar q)^{J=1}_{L,T} p}
(\vec l^{~\prime 2}_\bot, W^2) \left(1 - e^{-i \vec l^{~\prime}_\bot
\cdot \vec r^{~\prime}_\bot}\right)\cr
&&  = \left\{ \begin{array}{l@{\quad ~\quad}l}
\int d^2 l^\prime_\bot \bar \sigma_{(q \bar q)^{J=1}_{L,T}p}
\left( \vec l^{~\prime 2}_\bot , W^2 \right), &{\rm for}~ r^\prime_\bot
\to \infty, \label{7}\\
r^{~\prime 2}_\bot \frac{\pi}{4} \int d \vec l^{~\prime 2}_\bot
\vec l^{~\prime 2}_\bot \bar \sigma_{(q \bar q)^{J=1}_{L,T} p}
(\vec l^{~\prime 2}_\bot, W^2), &{\rm for}~ r^\prime_\bot \to 0 ,
\end{array} \right.
\end{eqnarray}
where $\vec l^{~\prime} = \vec l_\bot / \sqrt{z(1-z)}$, and $\vec l_\bot$ is 
the transverse momentum of the absorbed gluon.

The color-dipole cross section becomes $r^\prime_\bot$-independent 
for $r^\prime_\bot$ sufficiently large (``saturation''). It vanishes, as 
$r^{\prime 2}_\bot$, for $r^\prime_\bot$ sufficiently small 
(``color transparency''). 
Note that the scale for the $r^\prime_\bot$ dependence is 
$W$-dependent. It is determined by the magnitude of the $\vec l^{~\prime
  2}_\bot$-moment of the dipole cross section in the third line of 
(\ref{7}).

An important conclusion on the ratio
\begin{equation}
R (W^2, Q^2) \equiv \frac{\sigma_{\gamma^*_L p} (W^2, Q^2)}{\sigma_
{\gamma^*_T p} (W^2 , Q^2)} 
\label{8}
\end{equation}
follows immediately from (\ref{5}), (\ref{6}) and (\ref{7}). Replacing
the transverse dipole cross section in (\ref{5}) by (\ref{6}), and 
noting that for suffiently large $Q^2$ and appropriate energy, $W$, the 
integral in (\ref{5}) is 
determined by the $r^{\prime 2}_\bot \rightarrow 0$ behavior of 
(\ref{7}), we obtain \cite{4}
\begin{equation}
R(W^2 , Q^2) = \frac{1}{\rho} \frac{\int d^2 r^\prime_\bot r^{\prime 2}
_\bot K^2_0 (r^\prime_\bot Q)}{\int d^2 r^\prime_\bot r^{\prime 2}_\bot 
K^2_1 (r^\prime_\bot Q)} = \frac{1}{2\rho} , 
\label{9}
\end{equation}
where the mathematical identity
\begin{equation}
\int^\infty_0 dy y^3 K^2_0 (y) = \frac{1}{2} \int^\infty_0 dy y^3
K^2_1 (y)
\label{10}
\end{equation}
was inserted.

We note that a suppression of the longitudinal relative to the transverse
photoabsorption cross section by the factor 0.5 in (\ref{9}) is due to the photon
wave function, more precisely to the first moment of the photon wave 
function as a function of $r^\prime_\bot$ that enters as a 
consequence of color transparency. For $\rho = 1$ in (\ref{6}), i.e. 
helicity independence of the interaction of the $(q \bar q)^{J=1}$ 
fluctuation with the proton, $R (W^2, Q^2) = 0.5$. Any deviation from 
this value must be due to a helicity dependence of the $(q \bar q)^{J=1}p$
cross section, i.e. a dependence on whether the $(q \bar q)^{J=1}$ fluctuation 
originates from a 
transversely or a longitudinally polarized photon. For the 
transverse-size enhancement (\ref{4}) we find $R (W^2 , Q^2) = 0.375$, 
i.e. 
\vspace{-2pt}
\begin{equation}
R(W^2, Q^2)   = 
\left\{ \begin{array}{l@{\quad ~\quad}l}
\frac{1}{2} = 0.5 , 
&{\rm for}~ \rho = 1, ~{\rm helicity~independence},  \label{11}\\
\frac{3}{8} = 0.375 , 
&{\rm for}~ \rho = \frac{4}{3}, ~ {\rm transverse \!\!-\!\!size~enhancement}.
\end{array} \right.
\end{equation}  
\vspace{-2pt}
In terms of the structure functions $F_L (x, Q^2)$ and $F_2(x,Q^2)$, we have 
\begin{equation}
F_L (x, Q^2) = \frac{1}{1 + 2\rho} F_2 (x , Q^2)  = 
\left\{ \begin{array}{l@{\quad ~\quad}l}
0.33 ~ F_2 (x , Q^2) , 
& (\rho = 1),  \label{12}\\
0.27 ~ F_2 (x , Q^2) , 
&(\rho = \frac{4}{3}). 
\end{array} \right.
\end{equation}
We add the remark that the equalities (\ref{11}) and (\ref{12}) 
require sufficiently large $Q^2$. Quantitatively, in terms of the 
low-$x_{bj}$ scaling variable $\eta(W^2, Q^2)$ \cite{7}, 
\begin{equation}
\eta (W^2, Q^2) \equiv \frac{Q^2 + m^2_0}{\Lambda^2_{\rm sat} (W^2)}
> 1
\label{13}
\end{equation}
is required, where $m^2_0 \simeq 0.14~ {\rm GeV}^2$ and 
\begin{equation}
\Lambda^2_{\rm sat}
(W^2) \sim \int d~\vec l^{~\prime 2}_\bot \vec l^{~\prime 2}_\bot 
~ \bar\sigma_{(q \bar q)^{J=1}_L = 1} 
(\vec l^{~\prime 2}_\bot , W^2) \sim (W^2)^{c_2} .
\label{14}
\end{equation}

As seen in figs. 2 and 3, the experimental data are consistent with 
a transverse-size enhancement in (\ref{12}).\\ 
\begin{wrapfigure}{l}{0.55\columnwidth}
\vspace{-28pt}
\centerline{\includegraphics[width=0.54\columnwidth]
{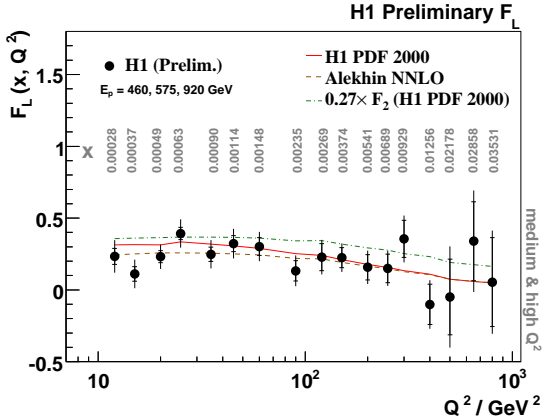}}
\vspace{-12pt}
\caption{\baselineskip=11pt
The prediction of $F_L (x, Q^2) = 0.27$ $F_2 (x, Q^2)$
compared with H1 experimental results 
(V. Chekelian, private communication).}\label{Fig:H1prelim}
\end{wrapfigure}  
The empirical validity of low-$x_{bj}$ scaling, $\sigma_{\gamma^* p} (W^2 ,
Q^2) = \sigma_{\gamma^* p} (\eta (W^2 , Q^2))$, was established \cite{7} in 
a model-independent analysis of the experimental data from HERA. Theoretically, 
low-$x_{bj}$ scaling is a consequence of the general structure of the 
color-dipole interaction (\ref{7}) combined with the (approximate) 
constancy of the $r^\prime_\bot \rightarrow \infty$ limit of the 
dipole-cross section in (\ref{7}), and dimensional analysis. 

For $\eta (W^2, Q^2) > 1$ or $Q^2 > \Lambda^2_{\rm sat} (W^2)$, where 
$2 {\rm GeV}^2 \le \Lambda^2_{\rm sat} (W^2) \le7 {\rm GeV}^2$ at HERA
energies,
both $F_2 (x,Q^2)$ and the gluon distribution 
$\alpha_s (Q^2) xg$ $(x, Q^2)$, 
using $x\equiv x_{bj}$, are proportional \cite{8} to the saturation scale,
$\Lambda^2_{\rm sat} (W^2)$, 
\begin{equation}
\left. \begin{array}{l@{\quad ~\quad}l}
F_2 (x,Q^2)  \\
\alpha_s (Q^2) xg (x,Q^2) & \cr 
\end{array} \right\} \sim \Lambda^2_{\rm sat} (W^2) \sim (W^2)^{c_2} . 
\label{15}
\end{equation}
\begin{wrapfigure}{r}{0.55\columnwidth}
\vspace{-18pt}
\centerline{\includegraphics[width=0.55 \columnwidth]
{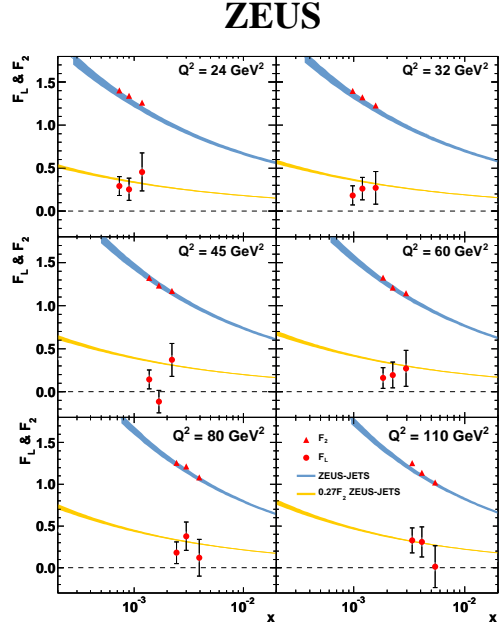}}
\vspace{-22pt}
\caption{\baselineskip=10pt 
As fig. 2, but compared with the ZEUS experimental results
(B. Reisert, private communication.) In case the originally yellow line $
F_L = 0.27 F_2$ is not well reproduced, compare the slides of
this presentation available under DIS2009.}\label{Fig:FLF2}
\end{wrapfigure} 
Consistency with DGLAP evolution \cite{9}, 
\begin{equation}
\frac{\partial F_2 \left( \frac{x}{2}, Q^2\right)}{\partial \ln Q^2} =
\frac{R_{e^+ e^-}}{9\pi} \alpha_s (Q^2) x g (x, Q^2) 
\label{16}
\end{equation}
requires \cite{8}
\begin{equation}
\frac{\partial}{\partial\ln W^2} \Lambda^2_{\rm sat} (2 W^2) = 
\frac{1}{2\rho + 1} \Lambda^2_{\rm sat} (W^2) 
\label{17}
\end{equation}
or 
\begin{equation}
(2\rho + 1) c_2 2^{c_2} = 1 . 
\label{18}
\end{equation}
Relation (\ref{17}) implies, respectively,
\begin{equation}
\rho = \left\{
\begin{array}{l@{\quad ~\quad}l}
1 , 
& c_2^{\rm theor.} = 0.276, \\
\frac{4}{3} , 
& c_2^{\rm theor.} = 0.23 . 
\end{array} \right.
\label{19}
\end{equation}
The result (\ref{19}) is consistent with the value from the model-independent
analysis of the experimental data \cite{7}, 
\begin{equation}
c_2^{\exp} |_{\rm Model-indep.} = 0.28 \pm 0.06 .
\label{20}
\end{equation}
Supplementing the CDP by the evolution constraint (\ref{18}) allows one to
predict $c_2$, i.e. the (strong) energy dependence, proportional to $(W^2)^{c_2}$
of $\sigma_{\gamma^* p} (W^2 , Q^2)$ and $F_2 (x,Q^2)$ for $Q^2 > 
\Lambda^2_{\rm sat} (W^2)$ in agreement with the experimental data. 

It is worth noting that the consistency of the evolution constraint (\ref{18})
on $c_2^{\rm theor.}$ with the experimental value of $c_2^{\exp}$ rules out
values of $\rho \gg 1$, as well as $\rho \ll \frac{4}{3}$, compare Table 1. 
The experimental result for the longitudinal-to-transverse ratio $R=1/2\rho
\simeq 0.375$ is indeed intimately related to the constant $c_2$ that
determines
the energy dependence of $\sigma_{\gamma^* p} (W^2, Q^2)$ and of $F_2 (x,
Q^2)$.
\begin{wraptable}{r}{0.45\columnwidth}
\vspace{-10pt}
\centerline{\begin{tabular}{|c|c|c|c|}
\hline
$\rho$  & $c_2^{\rm theor.}$ & $\frac{\sigma_{\gamma^*_L}}{\sigma_{\gamma^*_T}}$ & 
$F_2 \left( W^2 = \frac{Q^2}{x} \right) $ \\\hline
$\rightarrow \infty$  & 0 & 0 & $\left(\frac{Q^2}{x} \right)^0 =$ const 
\\\hline
0 & 0.65 & $\infty$ & $\left( \frac{Q^2}{x}\right)^{0.65}$  \\ \hline
\end{tabular}}
\vspace{-6pt}
\caption{The results for $c_2^{\rm theor.}$ for the assumptions of a 
very large and a very small value of $\rho$.}\label{Fig:table1}
\vspace{-40pt}
\end{wraptable} 

Since $c_2$ is correctly predicted by requiring (\ref{16}) to be valid for the 
structure function $F_2 (x,Q^2)=(Q^2  / 4\pi^2\alpha) \sigma_{\gamma^* p} (
\eta (W^2 , Q^2))$ in the CDP, the experimentally observed low-$x$ scaling does
not require non-linear effects in the evolution equations, neither for 
$\eta (W^2, Q^2) > 1$, nor for $\eta (W^2 , Q^2) < 1$. The saturation phenomenon 
for $\eta (W^2, Q^2) < 1$, where $\sigma_{\gamma^* p} (\eta (W^2, Q^2)) \sim
\ln (1 / \eta (W^2 , Q^2))$, is a consequence of the dipole interaction 
(\ref{7}). For sufficiently large energy, 
$\Lambda^2_{\rm sat} (W^2) \gg Q^2$, for any fixed $Q^2$, the 
photoabsorption cross section is determined by the $r^\prime_\bot \rightarrow 
\infty$ limit of the dipole cross section in (\ref{7}). 
For $\Lambda^2_{\rm sat} (W^2) \ll Q^2$, the color-dipole state interacts as a
dipole of vanishing size, $r_\bot\rightarrow 0$, 
while for $\Lambda^2_{\rm sat} (W^2) \gg Q^2$, it interacts as an
ordinary hadron with the gluons in the nucleon.

Consistency of linear evolution and scaling at low $x$ has recently also been
found \cite{10} by examining the double-asymptotic scaling approximation of 
the DGLAP evolution equations. 

\vspace{0.8cm}\noindent
{\bf Acknowledgments}

My thanks to Kuroda-san for a fruitful collaboration.

\end{document}